\def\sii{S~{\sc ii}}

\def\hi{H~{\sc i}}

\def\civ{C~{\sc iv}}
\def\feii{Fe~{\sc ii}}

\def\mgii{Mg~{\sc ii}}

\def\siii{Si~{\sc ii}}

\def\siiv{Si~{\sc iv}}

\def\alii{Al~{\sc ii}}

\def\znii{Zn~{\sc ii}}
\def\crii{Cr~{\sc ii}}

\def\oi{O~{\sc i}}

\def\ni{N~{\sc i}}

\documentclass[final,3p,twocolumn,authoryear]{elsarticle}
\usepackage{amssymb}
\usepackage[colorlinks=true,linkcolor=black, citecolor=blue, urlcolor=blue]{hyperref}
\usepackage{graphicx}
\usepackage{natbib}
\bibpunct{(}{)}{;}{a}{}{,}
\journal{New Astronomy}

\begin{document}
	\begin{frontmatter}
		
		%% Title, authors and addresses
		
		%% use the tnoteref command within \title for footnotes;
		%% use the tnotetext command for theassociated footnote;
		%% use the fnref command within \author or \address for footnotes;
		%% use the fntext command for theassociated footnote;
		%% use the corref command within \author for corresponding author footnotes;
		%% use the cortext command for theassociated footnote;
		%% use the ead command for the email address,
		%% and the form \ead[url] for the home page:
		%% \title{Title\tnoteref{label1}}
		%% \tnotetext[label1]{}
		%% \author{Name\corref{cor1}\fnref{label2}}
		%% \ead{email address}
		%% \ead[url]{home page}
		%% \fntext[label2]{}
		%% \cortext[cor1]{}
		%% \address{Address\fnref{label3}}
		%% \fntext[label3]{}
		
		\title{DLA and sub-DLA metallicity evolution: A case study of absorbers towards Q\,0338-0005}
		
		%% use optional labels to link authors explicitly to addresses:
		%% \author[label1,label2]{}
		%% \address[label1]{}
		%% \address[label2]{}
		
		\author[ist]{Waqas Bashir\corref{cor1}}
		\ead{chwaqasbashir@gmail.com}
		\author[aao]{Tayyaba Zafar}
		\author[ist]{Fazeel M. Khan}
		\author[wpa]{Farrukh Chishtie}
		\address[ist]{Institute of Space Technology, Institute of Space Technology, Islamabad, Pakistan.}
		\address[aao]{Australian Astronomical Observatory, PO Box 915, North Ryde, NSW 1670, Australia.}
		\address[wpa]{Department of Phsyics and Astronomy, Western University, London, Ontario, Canada, N6A 3K7}
		\cortext[cor1]{Corresponding author}
		\begin{abstract}
			The damped and sub-damped Ly$\alpha$ systems (DLAs and sub-DLAs) traced in absorption against bright background quasars represent the main reserve of neutral hydrogen at high redshifts. We used the archival Very Large Telescope (VLT) instrument Ultraviolet and Visual Echelle Spectrograph (UVES) high-resolution data of Q\,0338-0005 ($z_{\rm em}=3.049$) to study abundances of the DLA ($z_{\rm abs}=2.2298$) and sub-DLA ($z_{\rm abs}=2.7457$) along the line of sight. We estimated column densities of \hi\ and various elements present in the DLA and sub-DLA through Voigt profile fitting. The DLA throgh shows the Ly$\alpha$ emission from the its host galaxy. We derive the metallicities of the DLA and sub-DLA with [Zn/H] $=-0.67\pm0.18$ and [S/H] $=-1.45\pm0.17$, respectively. We compared our abundances of the DLA and sub-DLA with other high resolution DLA and sub-DLA metallicities and find that both populations show an overall increase of metallicity with decreasing redshift. However, sub-DLAs usually have higher metallicities than the DLAs.
		\end{abstract}
		
	\end{frontmatter}
	%% \linenumbers
	
	%% main text
	\section{Introduction}
	Quasar lines of sight provide a wealth of information about intergalactic medium and galaxies through the analysis of absorption line systems along with their sightlines. The detection of damped Lyman-$\alpha$ systems (DLAs; log N(\hi)$\geq20.3$ cm$^{-2}$) and sub-damped Ly$\alpha$ systems (sub-DLAs;  $19.0\leq$ log N(\hi) $\leq20.3$) against luminous background quasars does not depend on the brightness of associated galaxies rather on the cross-section of neutral hydrogen gas \citep[e.g.,][]{Wolfe1986,Wolfe2005}. These systems represent the main source of neutral hydrogen at higher redshifts \citep[e.g.,][]{Storrie-Lombardi2000,Noterdaeme2009,Tayyaba2013b}. Moreover, these are excellent laboratories, to study metals, dust, molecules, and to estimate the cosmological evolution of \hi\ gas mass density. 
	 
	Evolution and formation of galaxies is a key aspect in current observational cosmology and especially the gradual build-up of metallicity over cosmological time \citep[e.g.,][]{sommer08}. DLA surveys also reported systematic absorption from neutral and ionised species (\ni, \alii , \oi , \feii , \sii , \siiv , \civ\ etc.), demonstrating a complex interstellar structure observed in galaxies at present time \citep{Wolfe2000, Dessauges2003, Fox2007a, Fox2007b,zafar14b,zafar14}. To understand absorption gas properties, high resolution spectroscopy is crucial to better estimate the metal column densities. Sub-DLAs are found to have super-solar metallicities compared to the DLAs \citep{Kulkarni2007}. Therefore, it is important to compare both absorbers over redshifts to study their metal evolution to understand the metal budget at high redshifts. We here use the Very Large Telescope (VLT) Ultraviolet and Visual Echelle Spectrograph (UVES) data of quasar Q\,0338-0005 (RA: 03\,38\,54.78 Dec: $-00$\,05\,20.99; $z_{\rm em}=3.049$) reported in \citet{Tayyaba2013a} to study the DLA at $z_{\rm abs}=2.2298$ and a sub-DLA at $z_{\rm abs}=2.7457$ along its line of sight. The hydrogen column densities of both damped absorbers are published by \citet{Noterdaeme2009} with log N(\hi) $=21.05\pm0.25$ of the DLA and log N(\hi) $=20.17\pm0.47$ of sub-DLA. These column densities are derived from the low-resolution data from the Sloan Digital Sky Survey (SDSS) Data Release (DR) 12. In this work, we estimated metallicities for both systems and compared them with other DLAs and sub-DLAs from literature to study metallicity evolution.
	
	\section{Methods}
	
The high-resolution VLT/UVES \citep{Dekker2000} spectroscopic data for Q\,0338-0005 is taken under the programmes 074.A-0201(A) (PI: Srianand) and 080.A-0014(A) (PI: Lopez) with BLUE346, BLUE437, RED570, and RED760 settings. The wavelength coverage of the combined spectrum is 3030--9460\,\AA\ with a spectral resolution ($R=\lambda/\Delta\lambda$) of 90,000 and 110,000 for BLUE and RED arm, respectively. Standard UVES pipeline has been used to reduce data \citep{ballester00}. Individual spectra are merged and normalised within the ESO-\texttt{MIDAS} environment. Local continuum of the merged spectrum was determined by using spline function passing through the spectral regions and smoothly connecting the regions free from absorption lines. This is done to perform the column density analysis of the absorption lines. Because of multiple spectra obtained, the good signal-to-noise regions are 3760--4998\,\AA, 5680--7500\,\AA, and 7662--9320\,\AA.
	
	\begin{figure}
		\centering 
		{\includegraphics[width=\columnwidth,clip=]{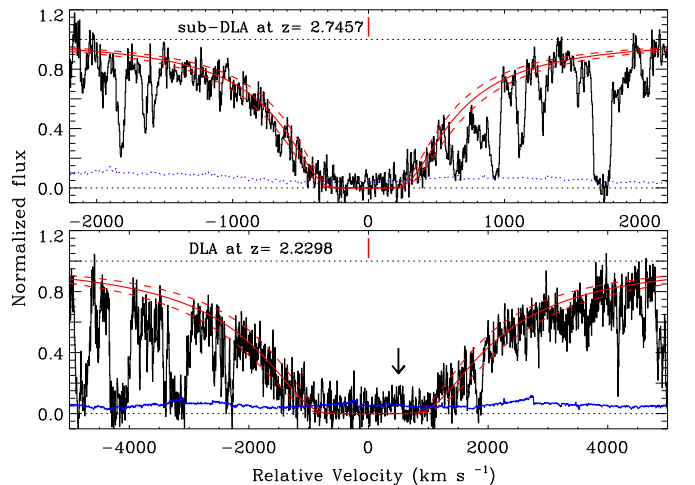} }
		\caption{{\it Top panel:} Combined 1D spectrum of the Q\,0338$-$0005 in the sub-DLA Ly$\alpha$ absorption region. The red solid and dashed lines show the best-fit \hi\ column density profile ($z=2.7457$, N(\hi) $=20.10\pm0.08$ cm$^{-2}$) and its 1$\sigma$ uncertainty, respectively. {\it Bottom panel:} DLA ($z=2.2298$) Ly$\alpha$ absorption region. The red solid and dashed lines show the best-fit \hi\ column density profile (N(\hi) $=21.09\pm0.10$ cm$^{-2}$) and its 1$\sigma$ uncertainty, respectively. The Ly$\alpha$ emission from the DLA host galaxy is marked by an arrow. In both panels, the blue dotted lines indicate the 1$\sigma$ error spectrum.}
		\label{lya}
	\end{figure}
	
	At redshift of DLA ($z_{\rm abs}=2.2298$) and sub-DLA ($z_{\rm abs}=2.7457$) we identify several metal absorption lines spread in the quasar spectrum. We identify lines from \oi\ ($\lambda 1302$\,\AA), \feii\ ($\lambda\lambda\lambda\lambda\lambda 2249, 2260, 2382, 2586, 2600$\,\AA), \siii\ ($\lambda\lambda\lambda 1304, 1526, 1808$\,\AA), \znii\ ($\lambda\lambda 2026, 2062$\,\AA), \crii\ ($\lambda\lambda\lambda 2026, 2056, 2062$\,\AA), and \mgii\ ($\lambda\lambda 2796, 2803$\,\AA) for the DLA which spread over twelve components. The lines from \sii\ ($\lambda\lambda 1253, 1259$\,\AA), \oi\ ($\lambda 1302$\,\AA), \feii\ ($\lambda\lambda\lambda 1608, 2374, 2382$\,\AA), \siii\ ($\lambda\lambda\lambda 1304, 1526, 1808$\,\AA), and \alii\ ($\lambda 1670$\,\AA) elements are identified for the sub-DLA spreading over four components.
	
	\begin{figure}
		\centering 
		{\includegraphics[width=\columnwidth,clip=]{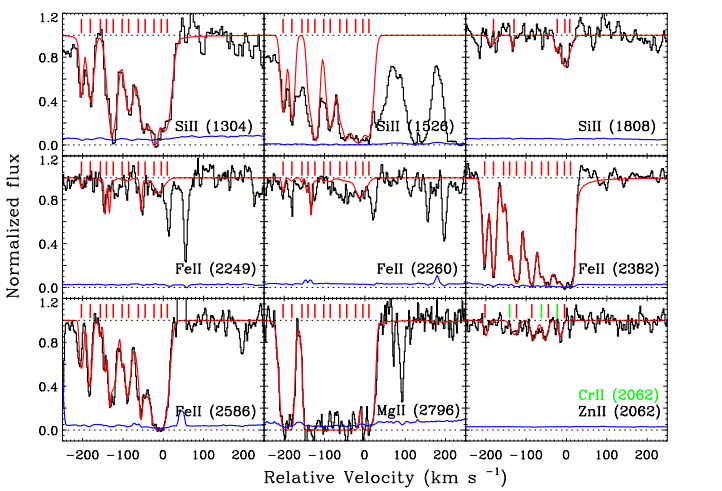} }
		\caption{Voigt-profile fits shown as red overlay to the metal absorption lines of the DLA at $z_{\rm abs}=2.2298$ towards Q\,0338-005. The vertical red line is our adopted zero velocity corresponding to $z_{\rm abs} = 2.2298$. Other redshift components are also indicated by red vertical lines. The green vertical lines indicate \crii\ absorption lines. The blue dotted lines show the 1$\sigma$ error on the spectrum. The horizontal dotted black lines indicate the levels at 0 and 1. }
		\label{fig:12}
	\end{figure}
	
We used the \texttt{MIDAS} Voigt profile fitting \texttt{FITLYMAN} package \citep{Fontana1995} to fit \hi\ column density and low-ionisation lines simultaneously. For the atomic data, laboratory wavelengths and oscillator strengths from \citep{Cashman17} were used. The \texttt{FITLYMAN} finds the best fit using a $\chi^2$ minimisation routine which fits the following function:
	
	\begin{equation}
	r(\lambda)=\frac{N r_0 f \sqrt{\pi} c \lambda_c 10^{-8}}{b\sqrt{2} \ln(2)}H(a,u)
	\end{equation}
	where Voigt function is defined by $H(a,u)$, $c$ is the speed of light, $f$ is the oscillator strength, $r_0$ is the classic radius of the electron, $a=\frac{\Gamma\lambda_c}{4\pi b 10^{13}}$, $u=\frac{(\lambda - \lambda_c)c}{\sqrt{2}\ln{(2)}b\lambda_c}$ with $\Gamma$ denoting damping coefficient and $b=\sqrt{b_K^2 + b_{turb}^2}$ with $b_K$ and $b_{turb}$ denoting Doppler thermal and turbulent broadening, respectively. The software includes the spectral resolution and returns best fit parameters for central wavelength ($\lambda_c$), column density ($N$), Doppler thermal and turbulent broadening (collectively as $b$) and 1$\sigma$ errors in each quantity.
	
\begin{figure}
		\centering 
		{\includegraphics[width=\columnwidth,clip=]{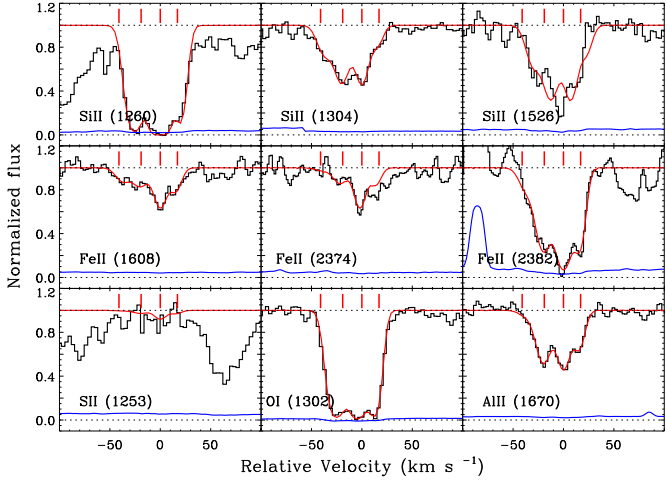} }
		\caption{Voigt-profile fits shown as red overlay to the metal absorption lines in the sub-DLA at $z_{\rm abs}=2.7457$ towards Q\,0338-005. The adopted zero velocity corresponding to $z_{\rm abs} = 2.7457$ is represented by vertical red line. For a comparison the other three redshift components are also indicated by red vertical lines. The vertical blue dotted lines represent the 1$\sigma$ error on the spectrum. The horizontal dotted black lines indicate the levels at 0 and 1.}
		\label{fig:13}
	\end{figure}

	Using the high-resolution data, we again fit the \hi\ column densities of damped absorbers again. We find the best-fit \hi\ column densities for DLA ($z=2.2298$) and sub-DLA ($z=2.7457$) are log N(\hi) $=21.09\pm0.10$ cm$^{-2}$ and log N(\hi) $=20.10\pm0.08$ cm$^{-2}$, respectively (Fig. \ref{lya}). Emission seen in the trough of the absorber is likely to correspond to the Ly$\alpha$ emission from the DLA host \citep[see e.g.,][]{kulkarni12,rahmani16,zafar17}. The Ly$\alpha$ emission from this DLA host galaxy is previously reported by \citet{krogager12} at impact parameters of $b=0.49\pm0.12$\,arcsec. The metal column densities of both QSO damped absorbers derived by fitting the normalised spectrum of the quasar with the \texttt{FITLYMAN} Voigt profile fitting routine. The best fit results are provided in Table \ref{dla} for the DLA and Table \ref{subdla} for the sub-DLA, respectively. The Tables provide for detected metals, the column densities in logarithmic form (log $N$) for each velocity component with a Doppler thermal and turbulent broadening ($b$). The resulting best fits in red are shown overlaid on the VLT/UVES data for both DLA and sub-DLA towards Q\,0338-0005 in Fig. \ref{fig:12} and Fig. \ref{fig:13}, respectively.
	
	\begin{table*}
		\centering
		\caption {Voigt-profile fitting of metal ion transitions in the DLA at $z_{\rm abs}=2.2298$ using \texttt{FITLYMAN}. Multiplets were fitted with same column densities plus turbulent broadening parameter values. The column densities of different metals in each velocity component are provided. The different transitions used for each metal are indicated in the second and third row.}
		\label{dla}
		\setlength{\tabcolsep}{1pt}
		\begin{tabular}{ccccccc}
			\hline
			$b$ & log N(\siii/cm$^2$) & log N(\feii/cm$^2$) & log N(\oi/cm$^2$) & log N(\mgii/cm$^2$) & log N(\znii/cm$^2$) & log N(\crii/cm$^2$) \\
			km~s$^{-1}$ & $\lambda\lambda$1304, 1526 & $\lambda\lambda$2249, 2260 & $\lambda$1302 & $\lambda\lambda$2796, 2803 & $\lambda\lambda$2026, 2062 & $\lambda\lambda$2026, 2056 \\
			& $\lambda$1808 & $\lambda\lambda$2382, 2586 & & & & $\lambda$2062 \\
			\hline
			\hline
			$6.3\pm0.1$ & $14.41\pm0.02$ & $13.03\pm0.02$ & $14.58\pm0.05$ & $>14.88$ & $12.31\pm0.06$ & $\cdots$ \\
			$6.4\pm0.1$ & $14.52\pm0.02$ & $13.20\pm0.02$ & $14.69\pm0.04$ & $>14.96$ & $\cdots$ & $\cdots$ \\
			$3.8\pm0.1$ & $14.06\pm0.01$ & $12.26\pm0.02$ & $14.51\pm0.05$ & $14.43\pm0.04$ & $\cdots$ & $\cdots$ \\
			$5.1\pm0.2$ & $14.21\pm0.01$ & $12.85\pm0.01$ & $14.85\pm0.03$ & $>14.75$ & $\cdots$ & $\cdots$ \\
			$10.0\pm1.1$ & $14.31\pm0.01$ & $13.52\pm0.01$ & $>15.89$ & $>14.87$ & $12.58\pm0.06$ & $12.43\pm0.06$ \\
			$3.6\pm0.1$ & $14.70\pm0.02$ & $14.09\pm0.03$ & $>16.31$ & $>15.24$ & $\cdots$ & $\cdots$ \\
			$9.2\pm0.4$ & $14.41\pm0.02$ & $13.54\pm0.01$ & $14.74\pm0.04$ & $14.65\pm0.04$ & $12.08\pm0.07$ \\
			$6.0\pm1.2$ & $14.23\pm0.02$ & $13.07\pm0.01$ & $>16.38$ & $>15.31$ & $\cdots$ & $\cdots$ \\
			$3.7\pm1.4$ & $14.66\pm0.01$ & $14.46\pm0.02$ & $>15.91$ & $>15.05$ & $11.97\pm0.08$ & $12.57\pm0.06$ \\
			$13.4\pm1.4$ & $14.11\pm0.02$ & $13.52\pm0.02$ & $14.86\pm0.04$ & $>15.42$ & $\cdots$  & $\cdots$ \\
			$6.7\pm0.8$ & $14.89\pm0.01$ & $14.52\pm0.02$ & $>16.65$ & $>15.77$ & $12.17\pm0.07$ & $12.64\pm0.07$ \\
			$9.9\pm1.4$ & $14.22\pm0.02$ & $14.02\pm0.02$ & $14.91\pm0.03$ & $14.76\pm0.03$ & $\cdots$  & $\cdots$ \\
			\hline			
		\end{tabular}
	\end{table*}
	
	\begin{table*}
		\centering
		\caption {Voigt-profile fitting of metal ion transitions in the sub-DLA (at $z_{\rm abs}=2.7457$) using \texttt{FITLYMAN}. The columns have same meaning as in Table~\ref{dla}. The different transitions used for each metal are indicated in the second row.}
		\label{subdla}		
		\setlength{\tabcolsep}{2pt}
		\begin{tabular}{cccccc}
			\hline
			$b$ & log N(\sii/cm$^2$) & log N(\siii/cm$^2$) & log N(\feii/cm$^2$) & log N(\oi/cm$^2$) & log N(\alii/cm$^2$) \\
			km~s$^{-1}$ & $\lambda\lambda$1253, 1259 & $\lambda\lambda\lambda$1260, 1304, 1526 & $\lambda\lambda\lambda$1608, 2374, 2382 & $\lambda$1302 & $\lambda$1670  \\
			\hline
			\hline
			$9.0\pm2.6$ & $12.40\pm0.11$ & $13.26\pm0.03$ & $12.97\pm0.15$ & $13.99\pm0.08$ & $11.05\pm0.23$ \\
			$9.0\pm0.1$ & $13.11\pm0.05$ & $13.64\pm0.03$ & $13.11\pm0.11$ & $14.59\pm0.04$ & $12.15\pm0.03$ \\
			$8.1\pm1.1$ & $13.52\pm0.06$ & $13.61\pm0.01$ & $13.46\pm0.06$ & $14.65\pm0.04$ & $12.18\pm0.05$ \\
			$6.4\pm1.5$ & $13.03\pm0.06$ & $12.92\pm0.03$ & $13.09\pm0.12$ & $14.43\pm0.05$ & $11.83\pm0.09$ \\
			\hline
		\end{tabular}
	\end{table*}
	
	The VLT/UVES resolving power is high enough to better resolve the lines within a few km/s. Weak lines, which are not saturated and lie on the optically thin region of the curve of growth, are used to extract accurate column densities. Strong lines have issues of saturation and because of this reason column densities derived from these are underestimated. Because of this reason, abundances of \oi\ and \mgii\ are considered to be lower limits. Individual element abundances in the neutral gas phase via comparison to the solar photospheric neighborhood \citep{Asplund2009} are provided for the DLA and sub-DLA in Table \ref{dlamet}. The hydrogen column densities derived from the UVES data are used to obtain the relative abundances.
	
	\begin{table}
		\centering
		\caption {Individual element abundances via analogy to the solar neighbourhood \citep{Asplund2009} for the DLA and sub-DLA towards Q\,0338-0005.}
		\label{dlamet}		
		\begin{tabular}{lccc}			
			\hline			
			Absorber & 		Ion &log \textbf{$N_{\rm tot}$}& [X/H] \\
			& & cm$^{-2}$ &  \\
			\hline
			\hline			
			DLA  & Ly$\alpha$ & $21.09\pm0.10$ & $\cdots$ \\
			& \siii\ & $15.52\pm0.07$ & $-1.08\pm0.12$ \\
			& \feii\ & $14.99\pm0.06$ & $-1.60\pm0.12$ \\
			& \mgii\ & $>16.26$ & $>-0.46$ \\
			& \oi\ & $>17.13$ & $>-0.65$ \\	
			& \znii\ & $12.98\pm0.15$ & $-0.67\pm0.18$ \\	
			& \crii\ & $13.03\pm0.11$ & $-1.70\pm0.15$ \\	
			\hline			
			sub-DLA & Ly$\alpha$ & $20.10\pm0.08$ & $\cdots$ \\
			& \sii\ & $13.77\pm0.15$ & $-1.45\pm0.17$ \\
			& \siii\ & $14.04\pm0.05$ & $-1.57\pm0.09$ \\
			& \feii\ & $13.80\pm0.06$ & $-1.80\pm0.10$ \\
			& \oi\ & $>15.12$ & $>-1.67$\\
			& \alii\ & $12.57\pm0.06$ & $-1.98\pm0.10$ \\
			\hline
		\end{tabular}
	\end{table}	
	\section{Discussion}
	The depletion of metal on to dust grain cause a fraction of refractory element (such as Fe, Si, Mg, and Al) to disappear from its gas phase. Si is marginally affected by dust depletion \citep{vladilo11} out of the gas phase but is not strongly depleted when compared to other refractory elements like Fe and Cr \citep{Ledoux2002, Draine2003, Fynbo2010, krogager16}. The best tracers of metallicity are zinc and sulphur. We looked for \znii\ absorption lines for both systems to have a reliable estimate of the metallicity. For sub-DLA, we cover only \znii\ ($\lambda$2062\,\AA) line between the good signal-to-noise regions, however, finding no significant detection. For DLA, we cover both transitional lines and detected \znii\ in 5 components (out of 12). We obtained Zn-based metallicity of DLA at $z_{\rm abs}=2.2298$ and used S-based metallicity for sub-DLA at $z_{\rm abs}=2.7457$ towards Q\,0338-0005. This gives us an opportunity to compare the metal content of these classes of objects and study their evolution over redshift. In the era of large quasar surveys, where thousands of DLAs are reported in the literature \citep{Noterdaeme2012}, sub-DLAs are not much explored and studied. Indeed, high spectral resolution and high signal-to-noise are required to estimate elemental abundances at lower \hi\ column densities of the sub-DLA. The studies of the high-resolution data available through spectrographs such as VLT/UVES are crucial to make our understanding better of the sub-DLAs and their properties.
	
\begin{figure}
		\centering 
		{\includegraphics[width=\columnwidth,clip=]{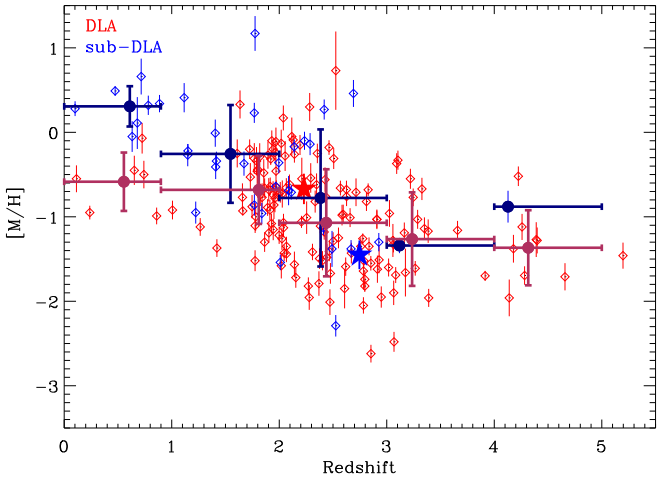} }
		\caption{Evolution of the mean metallicity [M/H] (where M is either Zn or S) of the damped absorbers with redshift. DLAs are shown in red colour while sub-DLAs are represented in blue colour. The DLA (red) and sub-DLA (blue) along the line of sight of Q\,0338-0005 are indicated as stars. The dark red and blue data indicate the binned metallicity of the DLAs and sub-DLAs respectively. Both DLA and sub-DLA populations show a trend of increasing metallicity with decreasing redshift. The sub-DLAs have higher metallicities compared to DLAs at all redshifts.}
		\label{met}
	\end{figure}

	DLAs plus sub-DLAs together contain the majority of neutral gas mass in the universe \citep{Tayyaba2013b}. Therefore, these classes of absorbers are an excellent tool to estimate the cosmic metal budget throughout the cosmic history. In Fig. \ref{met} we compared metallicities of the DLA and sub-DLA towards Q\,0338-0005 (shown as stars) with the other DLAs and sub-DLAs metallicities available from the literature. We here used both Zn and S abundances as a tracer of metallicity because of their less inclination to go onto dust grains.
	
	We collected Zn abundances for 122 and S abundances for 67 other DLAs and sub-DLAs from various high-resolution studies in the literature. In Fig. \ref{met}, the DLAs are shown in red colour and sub-DLAs in blue colour. The metallicities of both populations are compared with redshift to study the metal evolution. We binned the DLAs and sub-DLAs into five bins with a step-size of one in redshift. Central redshift for each bin, mean metallicity and standard deviation$+$ error (in quadrature) are plotted for both classes of damped absorbers. We have only one sub-DLA in each $3<z<4$ and $4<z<5$ bins, therefore, results are affected but low number statistics in those regions. Overall, the sub-DLAs have higher metallicities then DLAs. Although low statistics hinder to say much at high redshifts. Total sub-DLA metallicities reach near to the solar level at lower redshifts. Fig. \ref{met} also indicate that the total metal budget of both DLA and sub-DLA is overall decreasing at higher redshift which is consistent with the fact that at short timescales the absorbers are not evolved enough to produce enough metals in their interstellar medium.
	
	Previously, \citet{Kulkarni2007} consider a constant relative \hi\ gas in DLAs and sub-DLAs at low and high redshifts. According to them, the contribution of sub-DLAs to total metal budget increases with decreasing redshift. We find consistent results here where we further find that contribution of sub-DLAs to the metal budget is usually higher. \cite{Bouche2007} reported that at $z\sim2.5$, 17\% of the metals are in sub-DLAs but their estimate is highly dependent on the ionised fraction of the gas. It is important to compare sub-DLA metallicities with those of DLAs as done in this work. \citet{Som15} and \citet{Quiret16} also compared metallicity of sub-DLAs with the DLAs and find a steeper increase of sub-DLA metallicity with decreasing redshift than the DLA one.
	
	Not only observations present results of the total metal budget but also the chemical evolution models predict the metal budget. Models of cosmic chemical evolution claim that the global interstellar metallicity would decrease with increasing redshift such that the present day metallicity reaches the solar value \cite{Lanzetta1995, Pei1995, Pei1999, Tissera2001}. Our results and other studies suggest that sub-DLAs contribute substantially to the cosmic metal budget at all redshifts.
	
	\section{Conclusions}
	We estimated the column densities of various elements present in the DLA (at $z_{\rm abs}=2.2298$) and sub-DLA (at $z_{\rm abs}=2.7457$) along the line of sight of Q\,0338-0005 using the VLT/UVES spectrum. We derived the \hi\ column densities for both systems and detect Ly$\alpha$ emission from the DLA host galaxy. We calculated metallicity of the DLA with [Zn/H] $=-0.67\pm0.18$ and sub-DLA with [S/H] $=-1.45\pm0.17$. We put these damped absorbers in context with other high-resolution DLA and sub-DLA metallicities and find that both DLA and sub-DLA populations show an overall increase of metallicity with decreasing redshift. However, sub-DLAs have higher metallicities compared to the DLAs at all redshifts.
	
	\section{Acknowledgements}
	Based on the spectroscopic observations collected at the European Organisation for Astronomical Research in the Southern Hemisphere, 8.2 m Very Large Telescope (VLT) with the UVES instrument mounted at UT2 under ESO programmes 074.A-0201(A) and 080.A-0014(A).
	
	%% The Appendices part is started with the command \appendix;
	%% appendix sections are then done as normal sections
	%% \appendix
	
	%% \section{}
	%% \label{}
	
	%% If you have bibdatabase file and want bibtex to generate the
	%% bibitems, please use
	%%
	%%  \bibliographystyle{elsarticle-harv} 
	%%  \bibliography{<your bibdatabase>}
	
	%% else use the following coding to input the bibitems directly in the
	%% TeX file.
	
	\bibliographystyle{aa}
	\bibliography{q0338}
	
\end{document}